\documentclass{article}
\RequirePackage{filecontents}
\PassOptionsToPackage{dvipsnames}{xcolor}
\RequirePackage{comgipp}
\RequirePackage{amsfonts}
\RequirePackage{amsmath}
\RequirePackage{authblk}
  \newcommand{\theReferenceText}{\fullcite{Petrera2021}}

\usepackage[
sortcites,
backend=biber,
style=numeric,
firstinits=true,
url=false,
isbn=false,
safeinputenc,%
]{biblatex}

\usepackage{hyperref}
\usepackage{todonotes}
\usepackage{csquotes}
\usepackage{glossaries}
\newacronym{api}{API}{Application Programming Interface}

\newacronym{dlmf}{DLMF}{Digital Library of Mathematical Functions}

\usepackage{tabularx}

\providecommand{\fullTitle}{%
zbMATH Open: API Solutions and Research Challenges
}
\providecommand{\shortTitle}{%
zbMATH Open
}

\makeatletter
\@ifclassloaded{acmart}{%
  \title[\shortTitle]{\fullTitle}%
}{%
  \title{\fullTitle}%
  
}
\makeatother
\newcommand{\aff}[1]{\texorpdfstring{$^{#1}$}{}}
\author{Matteo Petrera\aff{1}, Dennis Trautwein\aff{2}, Isabel Beckenbach\aff{1}, Dariush Ehsani\aff{1}, Fabian Müller\aff{1}, Olaf Teschke\aff{1}, Bela Gipp\aff{2} and Moritz Schubotz\aff{1,2}}
 \affil{%
   $^1$ FIZ Karlsruhe, Berlin, Germany ({\{first.last\}@fiz-karlsruhe.de})\\
   $^2$ University of Wuppertal, Germany ({\{last\}@gipplab.org})\\
}

\begin{document}
  \maketitle
  \thispagestyle{firststyle}
  
  \begin{abstract}
We present zbMATH Open, the most comprehensive collection of reviews and bibliographic metadata of scholarly literature in mathematics.
Besides our website \url{zbMATH.org} which is openly accessible since the beginning of this year, we provide API endpoints to offer our data.
The API improves interoperability with others, i.e., digital libraries, and allows using our data for research purposes.
In this article, we
  (1) illustrate the current and future overview of the services offered by zbMATH;
  (2) present the initial version of the zbMATH links API;
  (3) analyze potentials and limitations of the links API based on the example of the NIST Digital Library of Mathematical Functions;
  (4) and finally, present the zbMATH Open dataset as a research resource and discuss connected open research problems.
  \end{abstract}
  
\section{Introduction}\label{sec:intro}

Since the beginning of 2021, zbMATH is open for public access. Currently, zbMATH Open contains over 4 million bibliographic entries with reviews or abstracts drawn from more than 3.000 journals and book series and more than 190.000 books. For most working mathematicians, this means that they can access zbMATH from anywhere in the world without subscription nor authentication. Additionally, we envision benefits to the community by our efforts to connect zbMATH data with information systems of research data, collaborative platforms, funding agencies, and intra-disciplinary efforts, as outlined in~\cite{Hulek20T,Schubotz21T}. We expect that our dissemination efforts of mathematics research results will increase the visibility of mathematics research. We invite the mathematical community to participate actively in the further development of the platform.

Very recently, at zbMATH, many efforts have been spent to develop \gls{api} solutions to facilitate and optimize open-access to mathematical research data. We expect that our efforts in disseminating mathematics research results will increase the visibility of mathematics for any scientific audience soon.

\begin{figure}
	\centering
	\includegraphics[width=0.98\columnwidth]{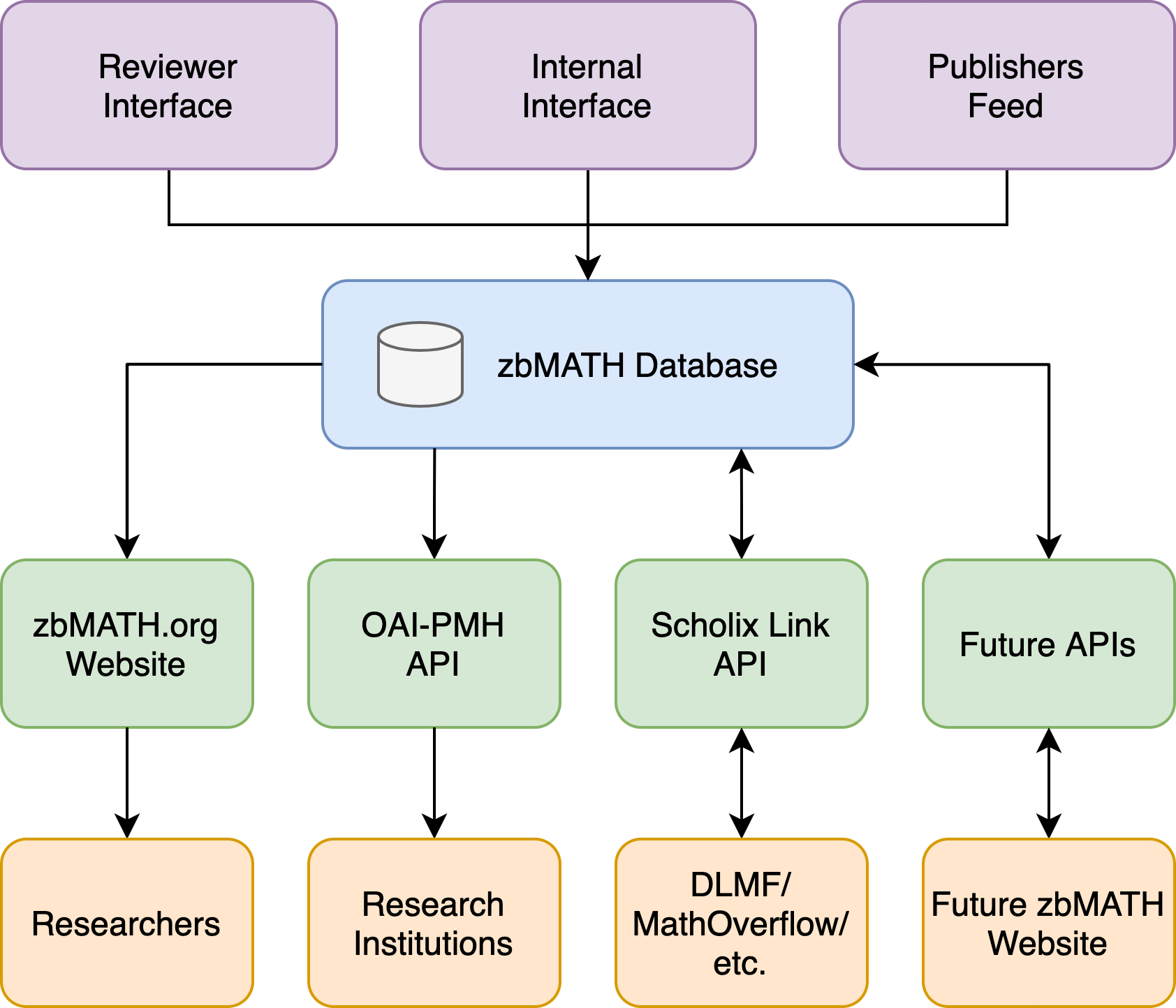}
	\caption{Conceptual overview of the zbMATH database and its associated ingress and egress data flows to various components. This paper puts the emphasis on the \enquote{Scholix Links API}.}
	\label{fig:1}
\end{figure}

In Figure \ref{fig:1}, we sketch a possible scenario for zbMATHs future APIs.
The boxes with \enquote{Reviewer Interface}, \enquote{Internal Interfaces}, and \enquote{zbMATH.org Website} show the well-established components of zbMATH. The green box \enquote{OAI-PMH API} was released in April 2021~\cite{Schubotz21T}.
This protocol is widely used for metadata-harvesting.
Via the OAI-PMH API, researchers can harvest the entire dataset or only specific subsets of our collection.
We offer the data in two flavors, the standardized Dublin Core\footnote{https://dublincore.org/} metadata format and a second format, that is closer to zbMATH's internal data model.
The content generated by zbMATH Open, such as reviews, classifications, software, or author disambiguation data are distributed under CC-BY-SA 4.0. This defines the license for the whole dataset, which also contains non-copyrighted bibliographic metadata and reference data derived from I4OSC (CC0). Note that the API does only provides a subset of the data in the zbMATH Open Web interface since in several cases third-party information, such as abstracts, cannot be made available under a suitable license through the API. In those cases we replaced the data with a placeholder string.
We envision that for researchers dealing with different data providers, the Dublin Core format is more suitable. We expect that for people used to our website, our own format is more appealing to use. From \newline
\centerline{\url{https://purl.org/zb/10}}\newline
one can fetch the entire dataset or a well-defined subset using a metadata harvester\footnote{https://www.openarchives.org/pmh/tools/}. One harvest output will be permanently stored as a research dataset of the Special Interest Group on Maths Linguistics data repository. This data repository also contains annual snapshots of arXiv articles in different formats optimized for mathematical information retrieval research challenges. As the zbMATH open data links to many arXiv preprints, we plan to synchronize the release cycles to create consistent snapshots of zbMATH data and associated fulltext sources.

In this paper, we describe a new service offered by zbMATH, namely an API, called \enquote{zbMATH Links API}, represented by the box stating \enquote{Scholix Link API} in Figure~\ref{fig:1}. At present, this new API is focused on the interconnections between zbMATH and the \gls{dlmf}\footnote{\url{https://dlmf.nist.gov/}}, even though more partners are expected to be hosted soon (e.g.,  MathOverflow, arXiv, Online Encyclopedia of Integer Sequences). Search engines or researchers from mathematics or the field of bibliometric research might use our zbMATH Links API to present and use the search results. Furthermore, the source code of our API has been released in the form of a Python package\footnote{\url{https://purl.org/zb/13}}, so that any interested user can use it for similar purposes in any context where the interconnection between bibliographic data and links has to be studied and documented.
In this way, we hope to serve the needs of a wide range of potential users.

The main contributions of this paper are:
\begin{enumerate}
    \item We provide a technical overview of the new API implementation using the example of how DLMF makes use of it. An analysis of the currently available dataset will be outlined.
    \item We present other natural candidates for the API, thus
    proving the potential coverage of the current mathematical literature.
	\item We highlight implications and new research potentials by showing how existing research can be transferred to make use of zbMATHs open APIs.
\end{enumerate}

In the following section \ref{sec:dlmf}, we motivate the choice of DLMF as the first partner for our new API and how it is currently used in their environment. Afterward, in section \ref{sec:linksapi}, we present the implementation details, analyze the DLMF link data and give some details about other potential partners. In section \ref{sec:method}, we discuss the technical capabilities of the new API and compare the capabilities of the open APIs of zbMATH with its pendant of PubMed. The last section is devoted to some concluding remarks and open problems.

\begin{figure}
	\centering
	\includegraphics[width=0.98\columnwidth]{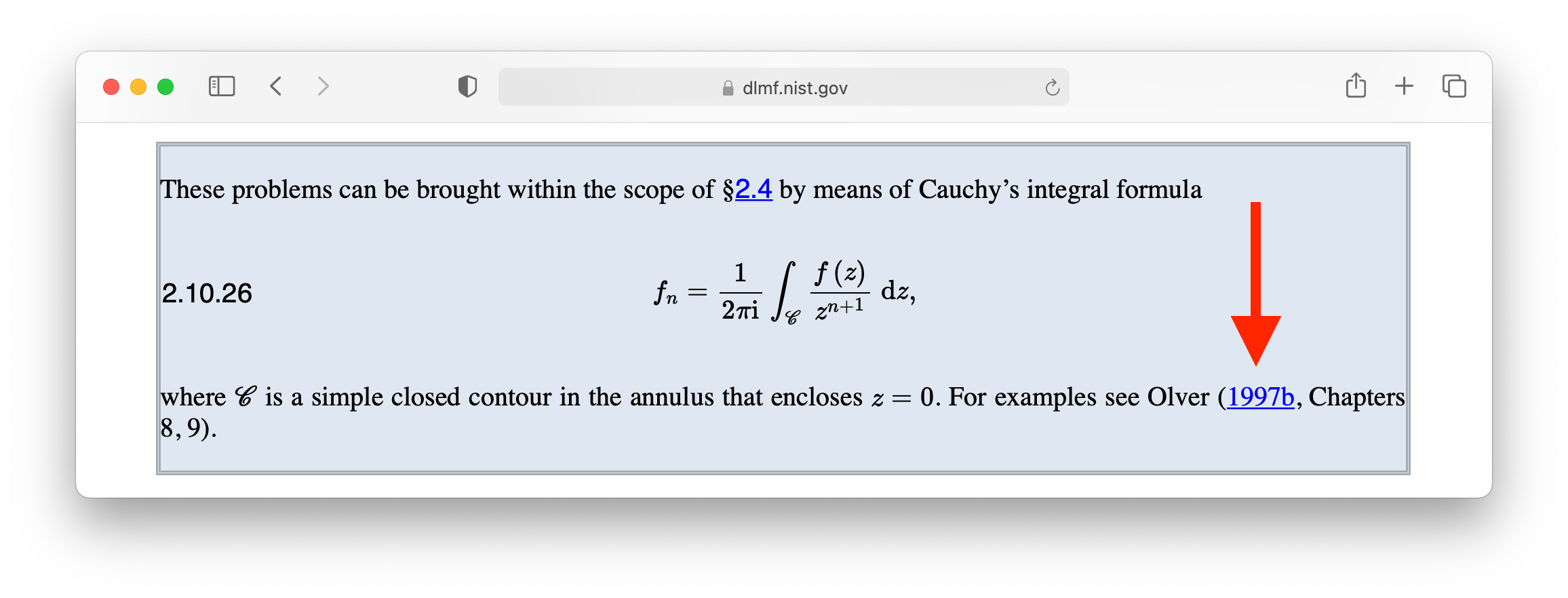}
	\includegraphics[width=0.98\columnwidth]{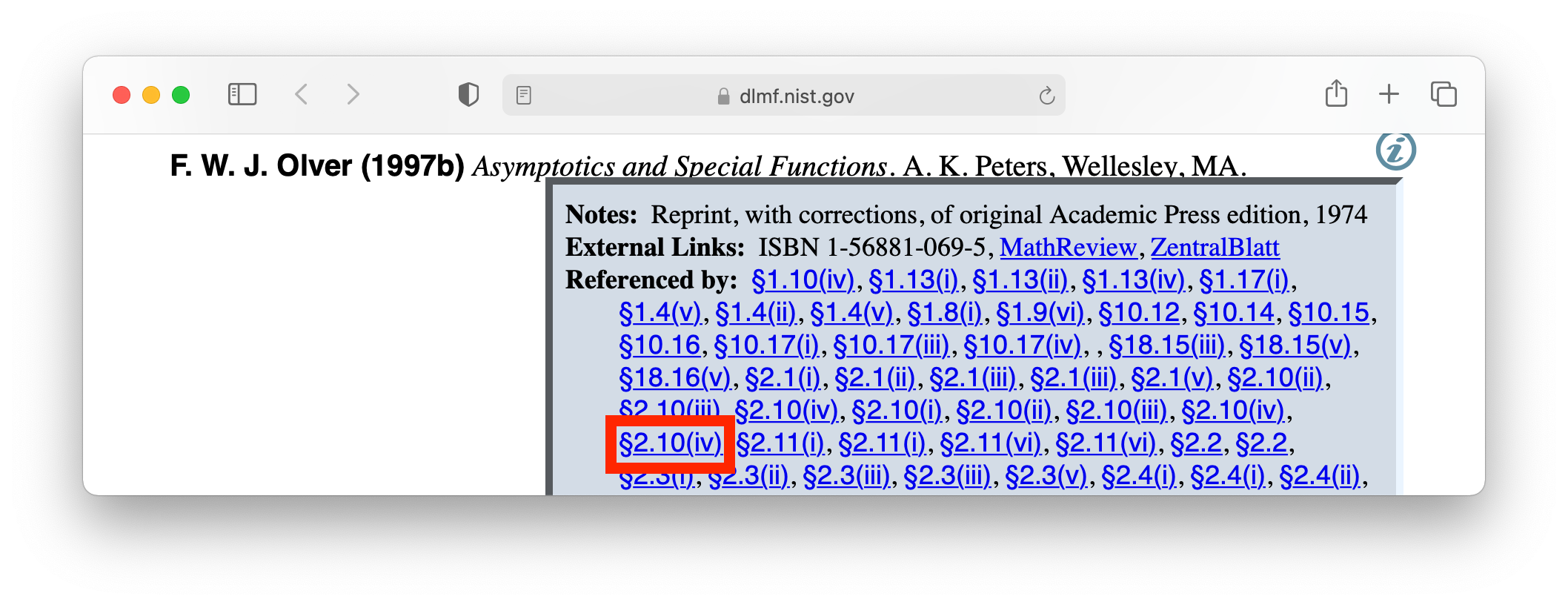}
	\caption{A reference in DLMF, available at \url{https://dlmf.nist.gov/bib/O} (below), and a link to it, \url{https://dlmf.nist.gov/2.10\#iv.p2} (above)}
	\label{fig:2}
\end{figure}

\section{DLMF as a zbMATH partner}\label{sec:dlmf}

Among all possible partners that may interact with zbMATH, we selected the aforementioned Digital Library of Mathematical Functions\footnote{\url{https://dlmf.nist.gov/}} as a first partner. In addition to being an important reference tool for mathematicians, DLMF offers a relatively small bibliographic catalog and is therefore very well suited for testing our API.

DLMF is a well-established web resource that enlarges and translates  the classical \enquote{Handbook of Mathematical Functions with Formulas, Graphs, and Mathematical Tables}, edited by M. Abramowitz and I. A. Stegun in 1964 into a modern and functional digital library.
As the original book's title inspiring this web service suggests, DLMF is a digital handbook about theoretical and computational aspects of special functions. 
Its primary purpose is to provide a modern reference tool for researchers in mathematics, physical sciences, and engineering.
It contains hundreds of definitions and theorems, presented with a standardized notation, together with tables, figures, and references to peer-reviewed papers and books.
It was published online on the May 7th 2010 and is continuously maintained, reviewed, and updated ever since.
Indeed, the field of special functions still receives great attention from the mathematics community, and new contributions enrich the contents of the library year by year.
DLMF presents its contents in 36 chapters, and the bibliography currently consists of 2.748 references\footnote{\url{https://dlmf.nist.gov/bib/}} of which 2.053 directly link to zbMATH (i.e., about 75\%).
This is  a valuable service offered independently by DLMF and zbMATH since each user has the possibility of accessing all selected publications' bibliographic data.
Let us note that of the remaining 25\% of publications not linked to zbMATH, most of them do not belong to the zbMATH database.

Before providing more details about our Links API, let us mention a few details about the links' structure we are interested in. Each reference in the DLMF bibliography may be cited many times in the DLMF pages.
Each of these instances carries its own link to zbMATH. For example, the book \enquote{Asymptotics and special functions} by F. W. J. Olver (Reprint, 1997; Zbl 0982.41018)\footnote{\url{https://zbmath.org/?q=an\%3A0982.41018}} is referenced 332 times.
Each citation defines a link to zbMATH uniquely. An example of one of these links is:
\url{https://dlmf.nist.gov/2.10\#iv.p2} (see Figure \ref{fig:2}).
In this case, Olver's book is referenced in Part 2 of Section \S2.10(iv) Taylor and Laurent Coefficients: Darboux’s Method.
In Figure \ref{fig:2}, we also see that the Section \S2.10(iv) is cited 3 times. Each instance corresponds to a link that points to a different destination site in the DLMF library. The highlighted \S2.10(iv) points to what we see in the first screenshot of Figure \ref{fig:2}.

\section{zbMATH Links API}\label{sec:linksapi}

This section presents the main features of the new \enquote{zbMATH Links API} by explaining its structure and various technical capabilities. Then, we give an analysis of the link statistics associated with our DLMF collaboration.

\subsection{Structure of the API}

The API itself has been implemented in Python and is described using the OpenAPI Specification\footnote{\url{https://swagger.io/specification/}}; a language-agnostic interface description standard for APIs.
At present, it hosts only one partner, DLMF, but it will soon host other partners.
The underlying dataset has been generated by scraping the DLMF bibliography.
As a result, we got 2.053 references (indexed at zbMATH) and 6.526 distinct links.
In this framework, the links are objects belonging to the \textit{source} (of a given partner; DLMF in the present case), and zbMATH objects are objects belonging to the \textit{target}.

The API offers eight endpoints, more specifically six \texttt{GET} routes, one \texttt{POST} route, and one \texttt{PUT} route. The  Swagger UI of the current zbMATH Links API is available at \url{https://purl.org/zb/14}. Here is a concise listing of the provided functionalities:

\begin{figure}
	\centering
	\includegraphics[width=0.98\columnwidth]{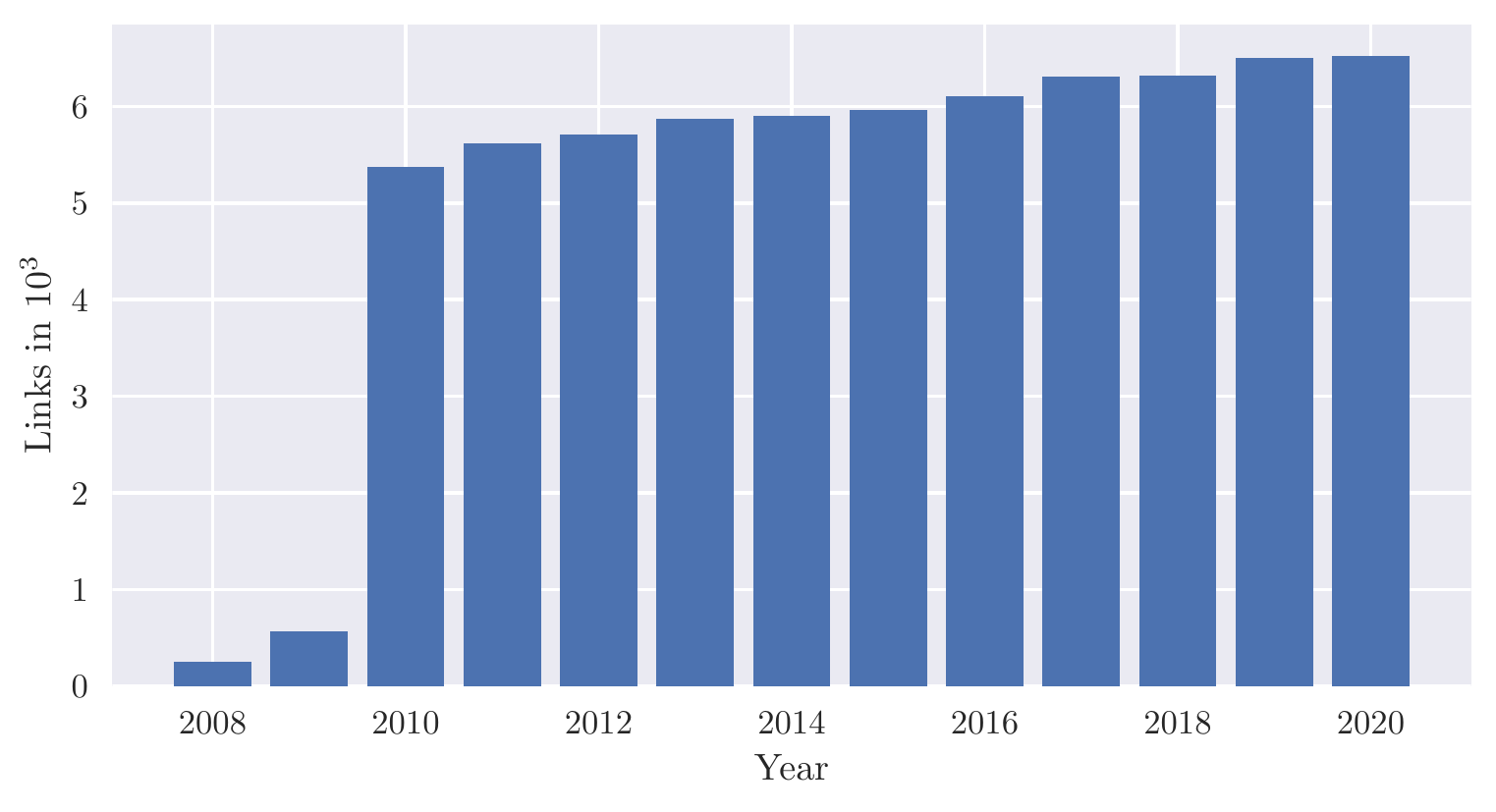}
	\caption{Number of links to the zbMATH API. One can see a huge increase in 2010 -- the year DLMF officially started.}
	\label{fig:backlinks1}
\end{figure}

\begin{itemize}
    \item \texttt{GET /partner} allows one to list all partners of zbMATH.
    \item \texttt{PUT /partner} allows one to edit a selected partner.
    \item \texttt{GET /link} allows one to retrieve links for given zbMATH objects. The parameters are: Authors, MSC codes\footnote{ Mathematics Subject Classification Scheme 2020, \url{https://msc2020.org/}}, X-Field\footnote{The X-Field is an optional parameter that can be used when one is running a query that can pull back a lot of metadata, but only a few fields in the output are of interest. Example: in the GET/link one is interested only in retrieving the id identifier of sources where the name of the author is Abramowitz. Then, Author: Abramowitz, X-Field: \(\{\)Source\(\{\)Identifier\(\{\)ID\(\}\)\(\}\)\(\}\).}.
    \item \texttt{GET /link/item} checks relations (if any) between a given link identifier (e.g., 2.10\#iv.p2) and a given zbMATH object (e.g., Zbl 0982.41018). The parameters are: Zbl code, Source identifier, Partner name, X-Field
    \item \texttt{POST /link} is intended to create a new link (for a given partner) related to a zbMATH object. The parameters are: Zbl code, Source identifier, Partner name, Link relation. This route allows one to interact actively with the dataset under consideration.
    \item \texttt{GET /source} provides a list of all links in the source.
    \item \texttt{GET /statistics/msc} shows the occurrence of primary MSC codes (2-digit level) in the source.
    \item \texttt{GET /statistics/year}  shows the occurrence of years of publication of references in the source.
\end{itemize}

Our JSON response body is modeled on the Scholix metadata schema\footnote{\url{http://www.scholix.org/schema/3-0}}.
The models used to pack the data are explicitly reported in the API web interface.
It is worth recalling that Scholix is a well-established framework to exchange information between data and literature links.
The schema's architecture is designed to allow for bulk exchange of link information, which contains all necessary data to keep track of bibliographic parameters identifying scholarly links.

\subsection{Analysis of DLMF Data}\label{sec:linksapi:eval}

Based on our available DLMF dataset, it is possible to draw some interesting conclusions:

\begin{figure}
	\centering
	\includegraphics[width=0.98\columnwidth]{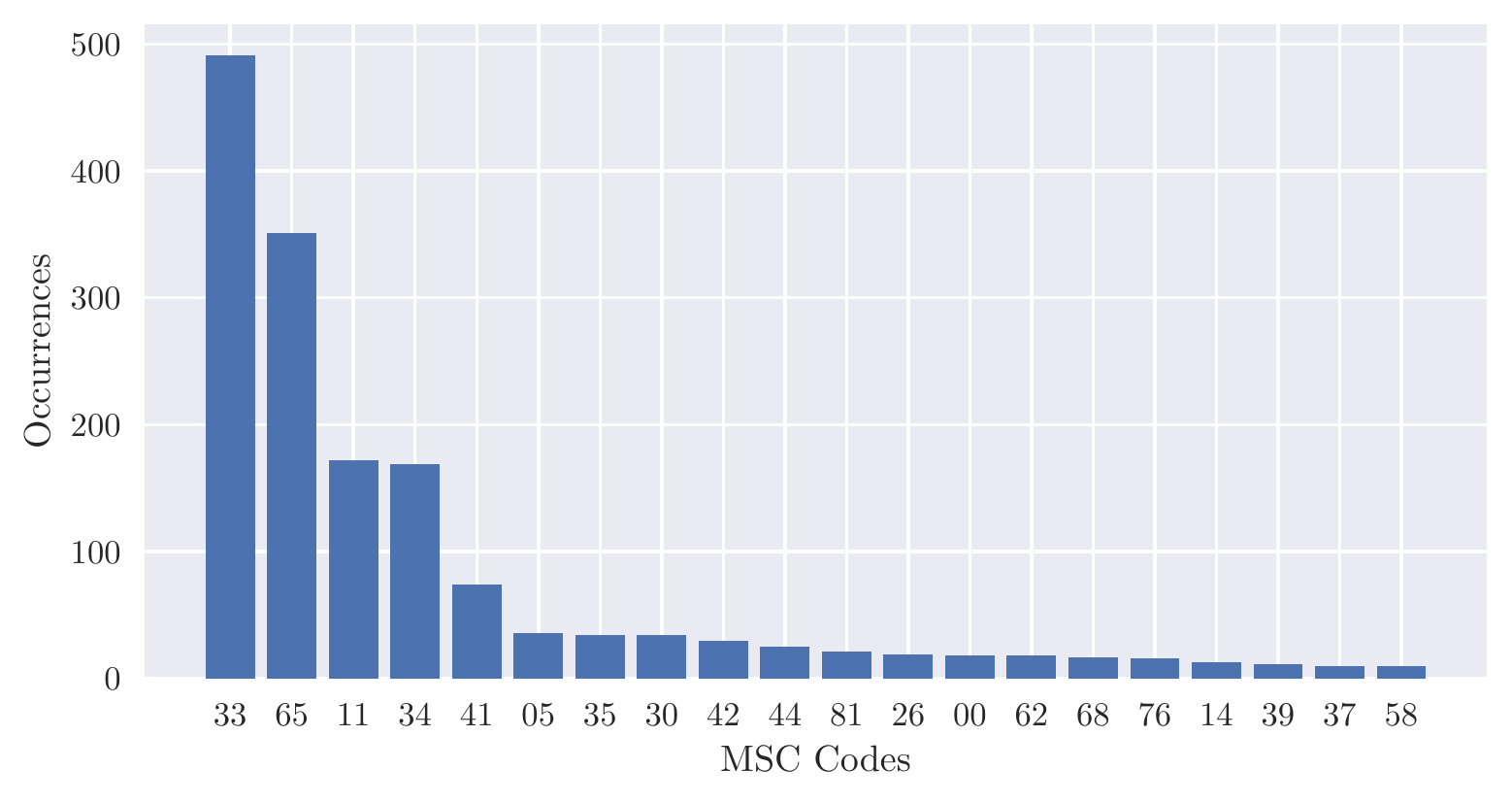}
	\caption{Distribution of primary 2-digit MSC codes in the DLMF dataset}
	\label{fig:backlinks2}
\end{figure}

\begin{figure}
	\centering
	\includegraphics[width=0.98\columnwidth]{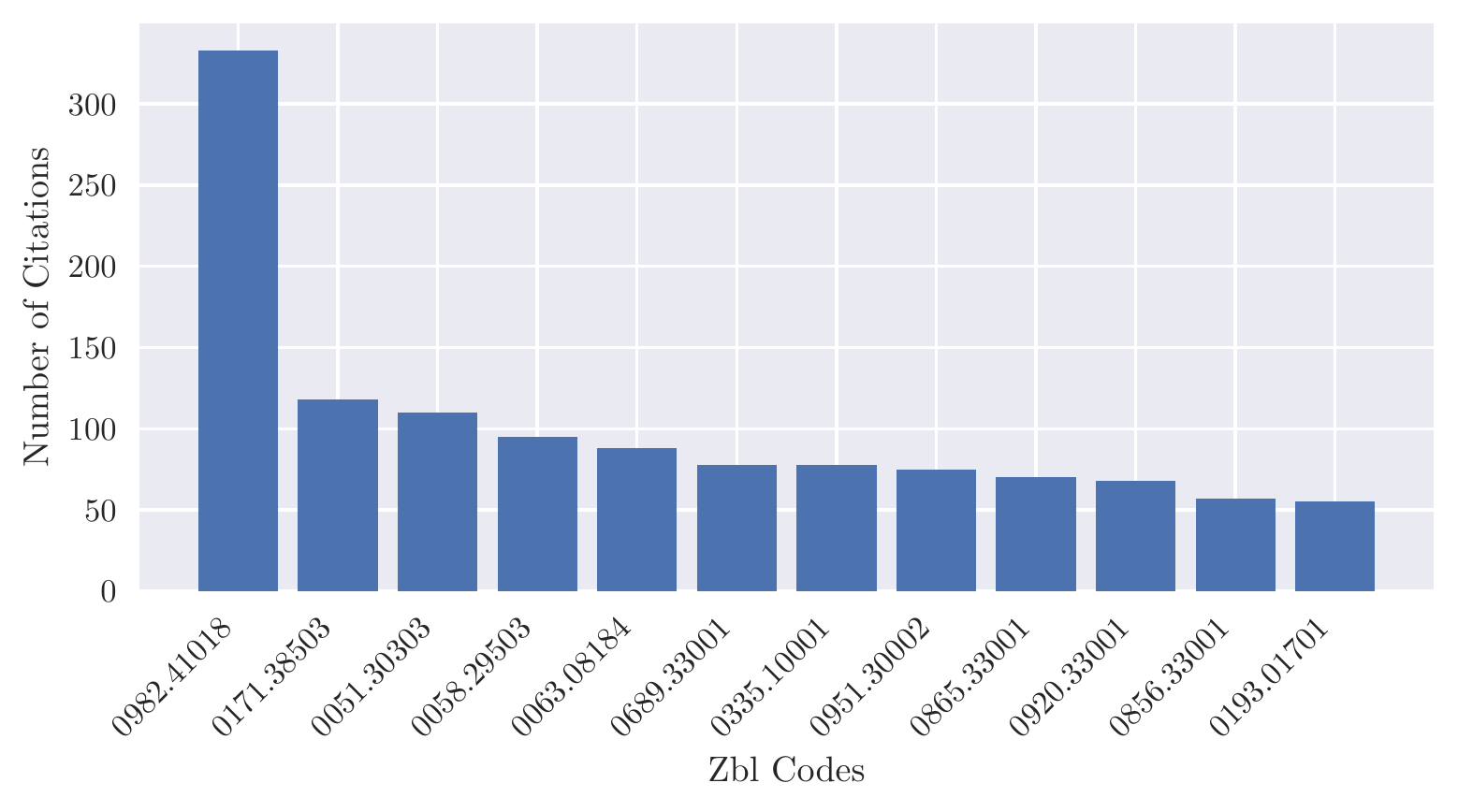}
	\caption{References (identified by Zbl code) in the DLMF dataset cited more than 50 times}
	\label{fig:backlinks3}
\end{figure}

\begin{itemize}
	\item In the JSON response body of our \texttt{GET /link} methods, one can see that each link is equipped with a publication date. This date refers to the date the link itself has been added in the DLMF bibliography. We scraped the historical bibliography between 2008 and 2021 (December is the scraping's reference month) and found the growth numbers depicted in Figure \ref{fig:backlinks1}. Clearly, the growth of population of references changed drastically in 2010, the year when DLMF started officially.
	
	\item The two statistics routes of our API show interesting results concerning the distribution of primary MSC codes (2-digit level) and years of publication of the references in the current dataset. As one may expect, the most frequently cited primary MSC codes are:

        \begin{table}[h]
            \center
            \normalsize
            \def\arraystretch{1.3}
            \begin{tabular}{ccl}
                \hline\hline 
                MSC Code & References & Area                \\ \hline
                33       & 491        & Special functions   \\
                65       & 351        & Numerical analysis  \\
                11       & 172        & Number theory       \\
                \hline\hline 
            \end{tabular}
        \end{table}

	      See Figure \ref{fig:backlinks2} for more details.
	      On the other hand the most frequent years of publication of cited references in the current dataset are:
            \begin{table}[h]
                \center
                \normalsize
                \def\arraystretch{1.3}
                \begin{tabular}{rccc}
                \hline\hline 
                     References & 67   & 65   & 65   \\ 
                     Year       & 1998 & 1999 & 1995 \\
                \hline\hline 
                \end{tabular}
            \end{table}
	      
	      See Figure \ref{fig:4} for more details. Looking at both
	      Figures \ref{fig:backlinks1} and \ref{fig:4} we could infer that the DLMF bibliography suffers from a delay in updating its references. More precisely, the fact that the maximum peak is centered at the end of the 90s makes us think of some kind of difficulty in identifying relevant references referring to the last twenty years. 
	          
	\item The references in the current DLMF dataset which have the most citations are:
	      \begin{itemize}
	      	\item \textit{F. W. J. Olver}, Asymptotics and special functions. Reprint. Wellesley, MA: A K Peters (1997; Zbl 0982.41018): 332 citations,
	      	\item \textit{M. Abramowitz} (ed.) and \textit{I. A. Stegun} (ed.), Handbook of mathematical functions with formulas, graphs and mathematical tables. Washington: U.S. Department of Commerce. (1964; Zbl 0171.38503): 118 citations,
	      	\item \textit{A. Erdélyi et al.}, Higher transcendental functions. Vol. I. New York: McGraw-Hill Book Co. (1953; Zbl 0051.30303): 110 citations.
	      \end{itemize}
	      In Figure \ref{fig:backlinks3} one can see the references, identified by Zbl code, with more than 50 citations. 
\end{itemize}

\subsection{Usage}
The motivation behind the recent implementation of APIs at zbMATH is twofold.
On the one hand, we want to offer the scientific community efficient and open-access to our data.
On the other hand, we wish to expose the dynamic interaction between our database's bibliographic data and those coming from other resources.
It is essential to note that both of these targets are made possible by zbMATH becoming an open web service.
This is a big incentive for disseminating scientific knowledge, and our work may help to understand how it spreads and auto-correlates in a very functional way.

The zbMATH links API with its first partner DLMF represents a tool that can be used in various ways and contains many properties that are advantageous for the research process. Here, we want to present concrete usage examples and examples of cases where a user of either DLMF or zbMATH can generally benefit from the service:

\begin{figure}
	\centering
	\includegraphics[width=\columnwidth]{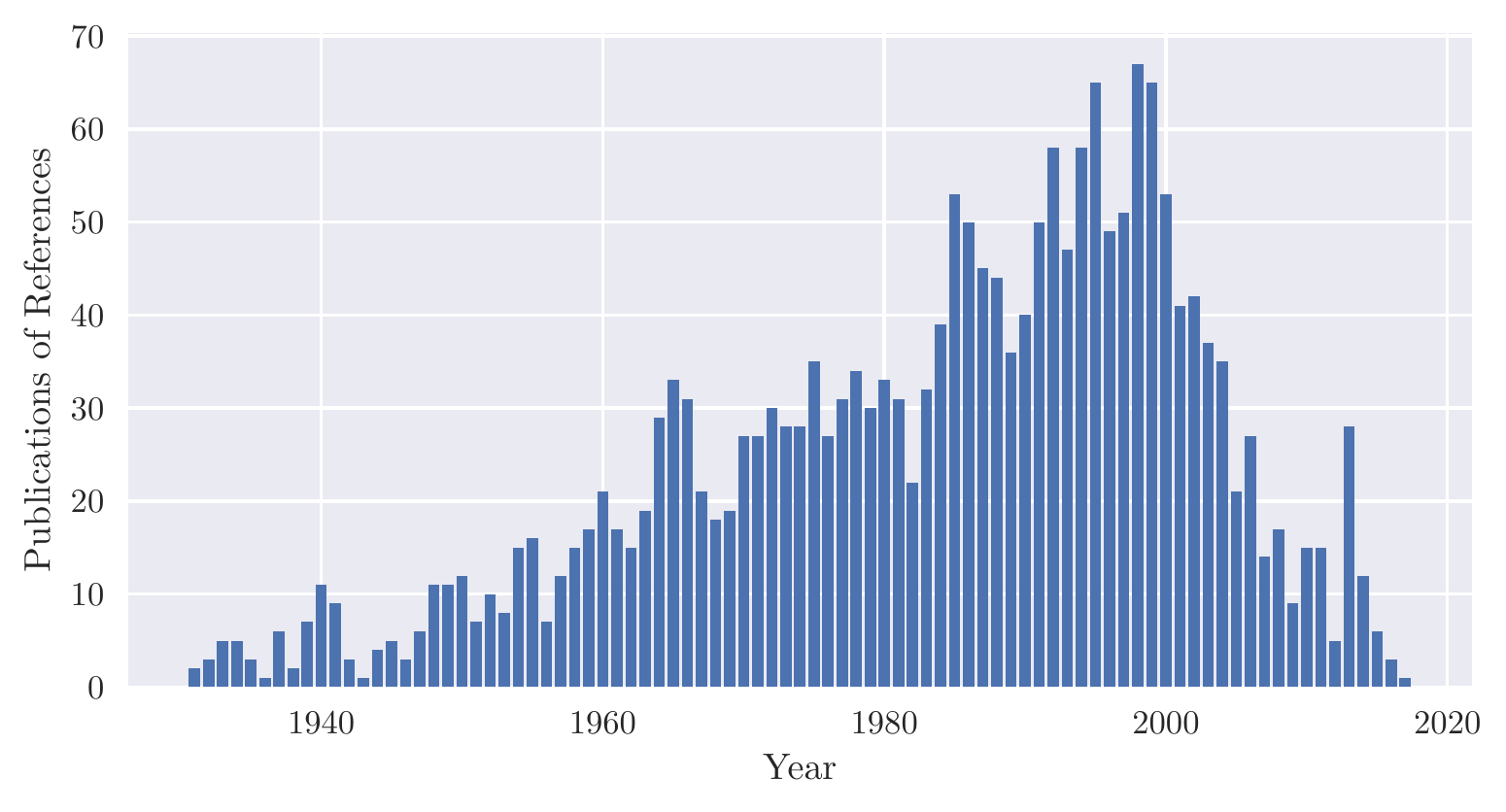}
	\caption{Distribution of years of publication of references in the DLMF dataset}
	\label{fig:4}
\end{figure}

\begin{itemize}

\item A DLMF user can access all bibliographic resources indexed at zbMATH relating to a specific topic of interest. This may help to get a consistent overview of the scientific development of the topic itself.

\item A researcher interested in a publication indexed at zbMATH can use our API to verify if and possibly where that publication is cited in DLMF. A search of this type can also be very diversified thanks to the filters that our routes offer. For example, one might be interested in identifying which DLMF links are related to a particular MSC code or a particular author. This means that targeted use of our API can allow a very detailed bibliographic search that otherwise would not be possible. 

\item A researcher more interested in the history of mathematics can use our API to trace the bibliography related to a certain topic covered in DLMF and observe the historical development of the topic itself in terms of the literature related to it. Such research can be very rich and diverse. It is sufficient to think that in the field of special functions there are classical topics, such as the \enquote{gamma function} or \enquote{elliptic integrals}, which have a long history behind them.
\end{itemize}

When other partners are included in our API, the covered spectrum will expand considerably, thus providing the user with an extremely efficient and flexible bibliographic searching tool.

In the following subsection, we will try to critically compare the functionality of our API solutions in relation to those offered by platforms similar to ours. Therefore, the goal is to understand in what aspects we can and must improve in the near future.

\subsection{Limitations and Future Partners}
While in general the Scholix API format, was a very good fit for our project we experienced some inconveniences.
For one, the link description in the DLMF sometimes contains mathematical expressions.
However, the API specification allows only string fields.
It would be good if the standard could be expanded to allow for HTML or another way of expressing mathematical expressions within descriptions.
Moreover, one of the problems we faced was modelling the MSC codes in the API.
We chose the field "subtype" of the "type" attribute in scholix.
However, this does not appear to be the original intent of that field.
Additionally, all MSC codes are joined to one string, which implies that those would be better modeled as an array, which is not allowed by the specification.
We are working on adding further partners to the zbMATH Links API. Three natural candidates are MathOverflow\footnote{\url{https://mathoverflow.net/}},  arXiv\footnote{\url{https://arxiv.org/}}, and the Online Encyclopedia of Integer Sequences\footnote{\url{https://oeis.org/}}.

\begin{itemize}
    \item MathOverflow is a question-and-answer platform for mathematics that is part of the StackExchange Network\footnote{\url{https://stackexchange.com/}}. In a previous collaboration, zbMATH and MathOverflow added the possibility to cite zbMATH entries in a MathOverflow post directly, see \cite{Mueller2019}. The zbMATH citations on mathoverflow.net link to the corresponding zbMATH record on zbmath.org. On the zbMATH side, we use the StackExchange API to generate links to MathOverflow questions citing a zbMATH record. This bidirectional linking is shown exemplarily in the two screenshots in Figure \ref{fig:zbmath_mathoverflow}. These data will soon be added to the zbMATH Links API.
    
    \begin{figure}
    	\centering
    	\includegraphics[width=0.98\columnwidth]{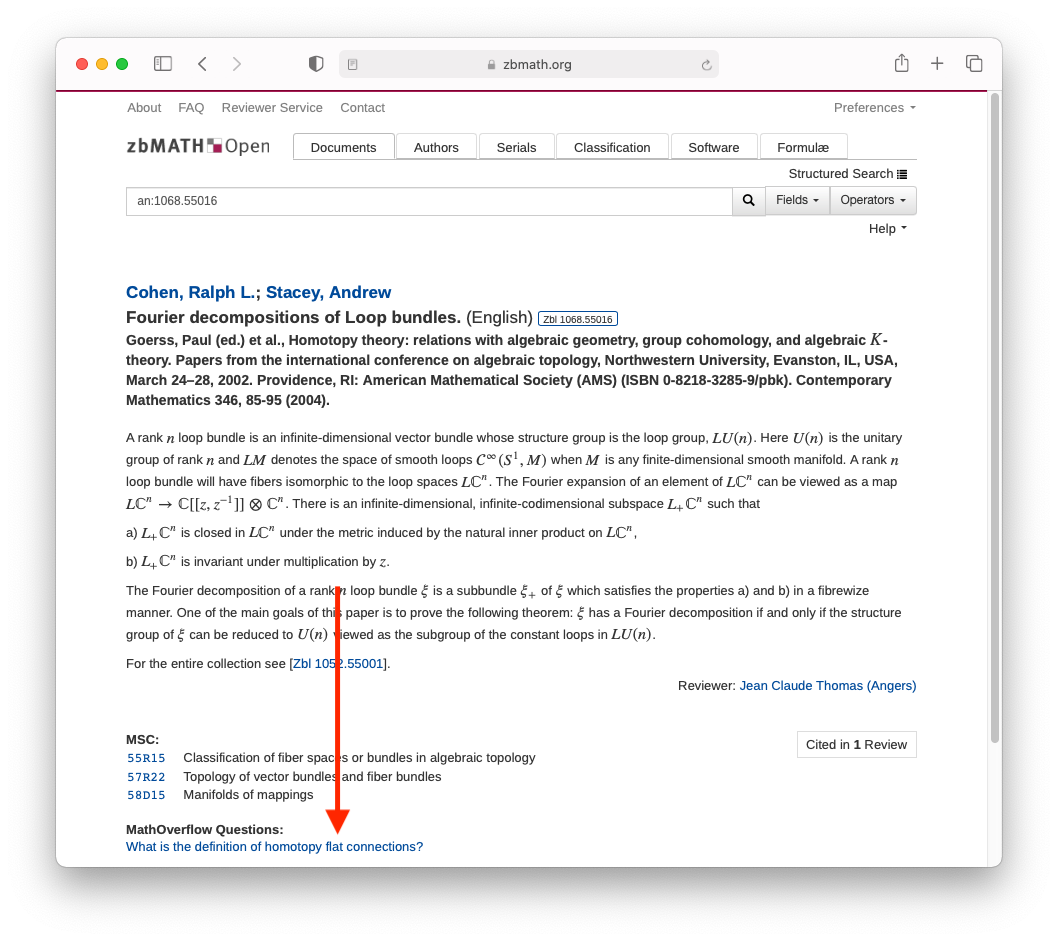}
    	\includegraphics[width=0.98\columnwidth]{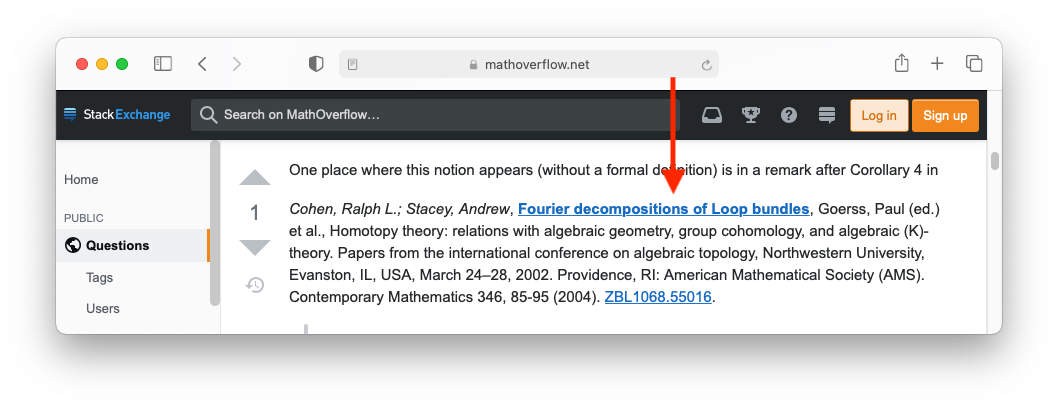}
    	\caption{A reference in MathOverFlow (below), and a link to it (above)}
    	\label{fig:zbmath_mathoverflow}
    \end{figure}

    \item arXiv is one of the most used open-access repositories of electronic preprints in mathematics. Roughly 250k zbMATH records contain links to their specific arXiv preprints that were added manually or thanks to information provided by the publishers. However, many arXiv preprints are still missing. To have access to an arXiv preprint of a zbMATH record is not only important for mathematicians, who might not have access to the journal version, but also to researchers who want to use the available arXiv data, which includes full-texts of many preprints, and combine this data with the metadata from zbMATH. Therefore, a suitable algorithm is needed to find a corresponding preprint for a zbMATH record if one exists. This problem can be seen as an entity matching problem, and there exists software for it, for example, JedAI, see \cite{Papadakis2018}. For our purpose, the existing software was not suitable as we wanted to use Elasticsearch\footnote{\url{https://www.elastic.co/elasticsearch/}} as a data source, and we have a lot of data (over 4 million zbMATH records, and over 800.000 arXiv preprints). Therefore, we implemented our own matching algorithms to find pairs of zbMATH records and arXiv preprints that correspond to each other.
    Our algorithm works differently than the usual entity matching algorithms. Instead of a blocking phase, it directly generates for each search record a small set of possible matching records (default is three), and compares each of these matching records with the search record. The possible matching records, which we will call candidates, are generated via an Elasticsearch query, where we search for the title and authors of a search record. To decide whether a search record and a candidate record match, a three-dimensional feature vector is computed. Then a trained decision tree classifier is used to map the feature vector to true or false. If multiple candidates match according to the classifier, we take the one whose feature vector has the smallest Euclidean norm. 

    To train the decision tree classifier and test our matching algorithm, we generated ground truth (or gold data) in the following way. We consider only arXiv preprints with a DOI in their metadata and search for a zbMATH entry with the same DOI. If we find one, we add this pair to our ground truth file. We also add some arXiv preprints with a DOI for which no zbMATH entry with the same DOI exists. We split the ground truth into a training set to train a decision tree classifier and a test set, which we use to evaluate our algorithm. We currently obtain a precision of 99.51 \% and a recall of 96.89 \% on the test set. However, given the large number of papers, we will need to improve the performance further to reduce the manual effort of correcting falsely linked papers.
    
    As for the previous example, the matching work done on arXiv will soon allow the addition of new links in our Links API.

    \item The Online Encyclopedia of Integer Sequences is a renowned online database of sequences of numbers launched in November 2010. It currently contains 342.422 sequences, each of them with its own list of metadata: first terms of the sequence, formulas for generating the sequence, references to books, articles, and scholarly links where the sequences have appeared, and more.
    At present, we are working on retrieving all references listed under \enquote{References} and \enquote{Links} for each sequence. Such references will be matched with our internal zbMATH Citation Matcher\footnote{\url{https://zbmath.org/citationmatching/}}
    and then stored in our Links API.

\end{itemize}

\section{Research Opportunities}\label{sec:method}
This section presents research opportunities connected to the newly released zbMATH Open data and API exemplified with concrete research questions.
Moreover, we compare our service with PubMed to put it in a broader context.
PubMed, with its underlying MEDLINE dataset and PubMed Central free full-text archive, is another well-known search engine within the biomedical scientific research and digital libraries community~\cite{DinhCP19, ErekhinskayaBTW16, EggersQSCM05, SaggionR17, PintoWB19, Veytsman19, BastK17, JhawarSSCBD20}.
It is available to the public since 1996, indexes over 32 million bibliographic references of biomedical literature, and is supported by the National Center for Biotechnology Information (NCBI), at the U.S. National Library of Medicine (NLM), located at the National Institutes of Health (NIH)\footnote{\url{https://pubmed.ncbi.nlm.nih.gov/about/}}. On the other hand zbMATH Open has over four million bibliographic entries and was made public on 1st January 2021. Table \ref{tab:comparison_zbmath_pubmed} shows a side-by-side comparison of PubMed and zbMATH Open.

\begin{table}
    \center
    \def\arraystretch{1.4}
    \caption{Side-by-side comparison of zbMATH Open and PubMed. These are the numbers from 2020}
    \label{tab:comparison_zbmath_pubmed}
    \begin{tabular*}{\columnwidth}{r@{\extracolsep{\fill}}ll}
        \hline\hline
                                       & zbMATH Open      & PubMed           \\ \hline
        Open Access since              & 2021             & 1996             \\
        Bibliographic Entries (Annual) & $>.13$ Million   & $> 1.5$ Million  \\
        Bibliographic Entries Total    & $> 4.0$ Million  & $> 31.5$ Million \\
        Journal Titles                 & $> 3.0$ Thousand & $> 5.0$ Thousand \\
        Search Queries 2020            & closed access    & $> 3.3$ Billion  \\
        \hline\hline 
    \end{tabular*}
\end{table}

We work out strengths and weaknesses by presenting selected research publications that leverage PubMeds APIs and analyze their applicability to the current state of zbMATH. This serves the purpose of uncovering immediate research opportunities in applying existing methods to the new open dataset of zbMATH and highlighting development prospects in areas where existing methods can not yet readily be applied due to missing interfaces or generally missing capabilities. The following paragraphs are to be understood as an inspiration for projects that can be based on the new open-access zbMATH data. After each paragraph, we propose one or multiple research questions that could follow from the described use case.%

\subsection{Immediate Research Opportunities}

In this subsection, we focus on research publications that have leveraged PubMeds open APIs and generally immediately research opportunities.

\subsubsection{Tagging of Scientific Publications}

Assigning keywords or tags to scientific publications is a crucial tool to increase discoverability. However, assigning such tags to scientific literature is an expensive and cumbersome process as human reviewers often assign them manually. This, in turn, leads to inconsistencies as different reviewers may assign different tags to the same publications. In \cite{Veytsman19} Veytsman proposes an automated approach to measure tag consistency across research publications based on a metric that captures how predictive a tag is for a citation. The author conducted experiments based on the MeSH\footnote{\url{https://www.nlm.nih.gov/bsd/disted/meshtutorial/introduction/index.html}} tags that human reviewers manually attach to documents of the PubMed database corpus. Each indexed publication of zbMATH contains one or many MSC codes\footnote{Mathematics Subject Classification 2020,\url{https://msc2020.org/}} and a set of keywords. The former is a hierarchical, alphanumerical identifier indicating the area of mathematics a certain research paper touches and the latter are free-text keywords that the authors suggest. Both classifiers are eventually adjusted by the editors of zbMATH.

We can imagine that the same experiments that Veytsman in \cite{Veytsman19} carried out can now be done based on the corpus of zbMATH Open. There would even be the possibility to further integrate with MathOverflow and recommend citations based on the tags given in their platform when a post is created.

Potential research questions:
\begin{enumerate}
\item \textbf{How to measure tagging consistency across mathematical research publications?}
Here, one can investigate how the methods developed in \cite{Veytsman19} can be applied to mathematics data. The required data can be derived via our API.
\item \textbf{What can be learned from crowd-sourced tagging in MathOverflow compared to curated tagging in zbMATH?} Especially interesting is here, if the tags from one service can help to for search in the other service.
The differences in the tagging behavior might also give insights on the learning curve as only known concepts will be tagged by individuals.

\end{enumerate}

\subsubsection{PDF Text Extraction Benchmark}

As the Portable Document Format (PDF) is the ubiquitous and standard format for scientific publications, its layout-based nature makes it hard to extract semantic meaning from the content. There exist a variety of tools that apply certain heuristics to identify which parts of a document represent, e.g., the title or a paragraph of text. Bast et al. \cite{BastK17} established a benchmark for text extraction performance of 14 tools by taking over 12.000 PDF documents from arXiv and obtaining their semantic information from associated \texttt{tex} files and then comparing the outputs of those tools to the semantic information present in the \texttt{tex} files. zbMATH Open also provides semantic information in the form of the XML format. While the investigated PDF files also contained some mathematics literature, the idiosyncrasies of mathematical typesetting may be worth a reevaluation with the sole focus on mathematics literature. Here especially the link between zbMATH entries and \texttt{tex} sources on arxiv which are provided by the API are helpful.

Furthermore, zbMATH Open provides high-resolution scans of early publications that were not yet typeset in a digital form alongside their corresponding \texttt{tex} source files for over 15.000 research article reviews. This corpus constitutes a huge potential for improving optical character recognition (OCR) techniques in the domain of mathematics as outlined in~\cite{Beck2020}.

Potential research questions:
\begin{enumerate}
\setcounter{enumi}{2}
\item \textbf{How do state-of-the-art PDF text extraction tools perform for mathematical literature?}
\item \textbf{What are the main challenges in optical character recognition of mathematical formulas?}
\end{enumerate}

\subsubsection{Training Dataset}

The opening up of zbMATH means that new training data can be used for artificial intelligence applications. The following listing provides inspiration for new possibilities that the dataset could be used for:

\paragraph{Formula Search}
The search mask of zbMATH Open already offers a formula search. However, the new open API allows building ones own or improving the formula search functionality by leveraging meta information provided alongside with the indexed articles.

Potential research questions:
\begin{enumerate}
\setcounter{enumi}{4}
    \item \textbf{What influence do different search options in digital libraries have on the scientific discovery process?} It is save to assume, that the discovery options for scientific literature will have an effect on the outcomes on ones own research. Here, one could try to qualitatively or even quantitatively assess this influence.
    \item \textbf{What are the state-of-the-art approaches to formula search, and what are the main challenges to overcome?}
\end{enumerate}

\paragraph{Recommender Systems} The provided data allows building a comprehensive recommendation system. This system could incorporate not only the meta information of the OAI-PMH APIs like MSC tags or keywords but also leverage the information on other platforms that a certain research article is linked in. E.g., mentions of related research papers in conversations on MathOverflow may be a good indicator for other relevant literature. As we continue to attract more and more partners for our Link API the context increases from which a potential recommender system can draw meaningful conclusions.

Potential research questions:
\begin{enumerate}
\setcounter{enumi}{6}
\item \textbf{Which features are most significant for related literature recommendations in mathematics?}
\item \textbf{What are the distinguishing challenges in feature extraction from mathematical literature?} The challenge of this research question is to identify how state-of-the-art recommender systems of other disciplines need to be tuned to excel at mathematical literature recommendations.
\end{enumerate}

\paragraph{Formula Disambiguation I} Similar formulas can have vastly different meanings in different contexts~\cite{DBLP:conf/sigir/ScharpfSG18, DBLP:conf/sigir/ScharpfSCG19, DBLP:conf/recsys/ScharpfMSBBG19, Scharpf2021}.
This is especially true for single symbols used in these formulas as researchers in different fields will certainly have assigned a different meaning to symbols. A system that tries to understand in which context a formula appears and draw meaning from that could especially leverage the MSC classification that is assigned to all articles on zbMATH Open. Most results from the OAI-PMH API contain an abstract where one can often find typeset formulas that can be used as training data along with full-text data that can be obtained through arXiv.

Potential research questions:
\begin{enumerate}
\setcounter{enumi}{9}
\item \textbf{How can similarly typeset formulas describing different concepts be disambiguated?} The main challenge of this research question is to devise criteria that make a formula ambiguous.
\item \textbf{What are the distinguishing factors in formula typesetting to avoid ambiguity?} In this research question it would be the goal to devise guidelines to avoid typesetting ambiguous formulas in the first place. 
\end{enumerate}

\paragraph{Formula Disambiguation II} Following the above disambiguation, it is also possible for a single concept to be expressed in different ways. Imagine the circumference $U$ of a circle being expressed in one paper as $U=2\pi r$ and in another $U=\pi d$ with radius $r$ and diameter $d$. Indeed, both formulas describe the same concept but are typeset differently. This kind of disambiguation will be of immediate relevance for academic plagiarism detection. State-of-the-art plagiarism detection systems already consider paraphrased text but lack capabilities to effectively detect \enquote{paraphrased} formulae~\cite{Meuschke2019}.

Potential research questions:
\begin{enumerate}
\setcounter{enumi}{11}
\item \textbf{How can differently typeset formulas describing the same concept be disambiguated?} The main challenge of this research question is to devise ways to identify such formula combinations.
\item \textbf{What factors make a formula more readable than a differently typeset formula describing the same concept?} Here, one can investigate factors for readability and if there are objectively better ways to typeset a certain formula.
\end{enumerate}

\paragraph{Math Spell-Checking} 
Popular tools like Grammarly\footnote{\url{https://grammarly.com/}} scan your text for common grammatical mistakes and provide the user hints about potential improvements. A similar offering could be developed for typesetting formulas by, for example, giving simple warnings of missing closed parentheses (if applicable) or other common mistakes. Such a spell-checking system could make use of the data of zbMATH Open and linked peripheral services. The linking to arXiv could be used to retrieve the full-text \texttt{tex} information, and the connection to MathOverflow could be used to detect common mistakes by taking into account the edit history of formulas in posts.

Potential research questions:
\begin{enumerate}
\setcounter{enumi}{13}
    \item \textbf{What are common errors in mathematical formula typesetting, and how to identify them?} The main challenge of this research question is to derive a method to identify erroneous formulas; and as a second step to investigate what common errors are.
    \item \textbf{What impact had formulas containing errors in the mathematics research community?} Here, one can research the consequences that errors in formulas and the research that built on them had. This could be extended to the influence of errors in formulas on widespread websites like Wikipedia to contemporary incidence.
\end{enumerate}

\paragraph{Classification and clustering}
While zbMATH Open provides MSC tags and keywords for the research articles, we can imagine that there are different classification and clustering approaches
that are not represented through the meta information of zbMATH. The open-access to the APIs allows building use case specific search and clustering systems.

Potential research questions:
\begin{enumerate}
\setcounter{enumi}{15}
\item \textbf{Do different logical classification and clustering schemes emerge from the zbMATH Open metadata besides the MSC classification scheme?}
\end{enumerate}

\paragraph{Review generation}
At present, many research papers and books indexed at zbMATH are supplemented with a review written by external experts in the field. Currently more than 7.000 active experts participate in compiling reviews for research papers and books. They critically analyze the contribution of the publication under consideration, often summarize the content and judge it in reference to a bigger context. With the advancements of text generating deep learning models such as GPT-3, it is not far to seek to train models on these handwritten reviews in conjunction with their full-text articles and metadata of zbMATH Open.

Potential research questions:
\begin{enumerate}
\setcounter{enumi}{16}
\item \textbf{What are significant properties a mathematical review should encompass?} In this research question one should distill the essential properties of what makes a \enquote{good} mathematical review.
\item \textbf{How do generated mathematical reviews with GPT-3 compare with manually written reviews according to the aforementioned significant properties?} Here, it is interesting if artificial intelligence is capable of meeting the aforementioned properties.
\item \textbf{What impact can artificial intelligence models such as GPT-3 have on the mathematical review process?} In this research question, one should work out the implications of potentially machine written reviews.
\end{enumerate}

\subsection{Development Prospects}
In this subsection, we focus on research publications that have leveraged PubMeds open APIs to which there is no pendant yet in zbMATH Open. The uses-cases in this section serve as inspiration for development opportunities.

\subsubsection{Retraction Tracking}\label{sec:method:retraction-tracking}

There are manifold reasons why a scientific publication could get retracted. It can range from erroneous study design to deliberate misconduct like plagiarism or generating artificial data to support a hypothesis. With the increasing amount of scientific literature at an accelerating rate, the number of retracted papers naturally increases as well. Therefore, it is crucial to notify researchers early in the research process about possible retracted publications. In \cite{DinhCP19} Dinh et al. present a Zotero\footnote{\url{https://www.zotero.org/}} plugin called \textit{ReTracker} that helps to identify retracted papers from PubMed. \textit{ReTracker} uses the full paper titles as they are present in the Zotero library to query PubMed on its retraction status. This status is persisted in a local cache and displayed to the user. With the opening of zbMATH this plugin could now not only cover articles of biomedical literature but to also inform researches about retracted publications in the field of mathematics. Currently, zbMATH Open does not provide information about the retraction status, but we can imagine that collecting this information from various trustworthy sources and making it accessible through the API would be a valuable addition to the current service. The authors in \cite{DinhCP19} underline the need for such a tool by stating that the citation rate of retracted publications can even increase after they got their retraction status \cite{DinhCP19}, so, literature is still cited even years after retraction.

Potential research questions:
\begin{enumerate}
\setcounter{enumi}{19}
\item \textbf{How does the retraction of mathematical papers influence their citations?} This question follows the observation of \cite{DinhCP19} that the citation count of literature still increases after it got retracted, so the intuitive answer that citations stop after retraction does not hold true. Here, it would be interesting to identify the reasons why literature is still cited.
\item \textbf{What are the most common reasons for the retraction of mathematical research papers, and how can publication of such papers be minimized?} Here, one can think in the direction of computer assisted quality assurance on the publisher side and how this could help the publishing process.
\end{enumerate}

\subsubsection{Collaboration Identification}

While digital libraries nowadays offer comprehensive and advanced search interfaces to retrieve and explore related scientific literature, they often lack the understanding of how authors have collaborated and to which extend their collaboration was fruitful. The same statement is true for zbMATH Open. In \cite{Cagliero2017IdentifyingCA} Cagliero et al. explored ways to identify collaboration patterns of authors and to measure to what extent the collaboration was fruitful. They harvested digital libraries and online databases for research publications and applied a pattern-based approach to identify collaborations among researchers. By making the APIs of zbMATH open-access, we believe that Cagliero et al. \cite{Cagliero2017IdentifyingCA} can serve as inspiration to motivate further insights generation techniques like author collaboration identification.

Potential research questions:
\begin{enumerate}
\setcounter{enumi}{21}
\item \textbf{How can the open data of zbMATH Open be used to construct collaboration graphs among mathematics researchers?} The main contribution in this research question would be a comprehensive collaboration graph based on the zbMATH open dataset.
\item \textbf{What conclusions can be drawn from an author collaboration graph concerning collaboration effectiveness?} Here, one can investigate how the methods developed in \cite{Cagliero2017IdentifyingCA} can be applied to the data of our API.
\end{enumerate}

\section{Conclusions and Future Work}\label{sec.concl}

In this article, we have presented the recent innovations made to zbMATH.
We implemented API solutions following the OAI-PMH and Scholix standards.
Those solutions allow the scientific community to use our open database in an efficient and reproducible way.
We demonstrated the capabilities of the API based on the links between DLMF and zbMATH.
By combining classification information from zbMATH with reference information from DLMF, we could derive new insights on references in the DLMF. 
In the future, we will incorporate MathOverflow, arXiv, and the Online Encyclopedia of Integer Sequences to the new zbMATH Links API. 
Moreover, we gave inspiration for research opportunities arising from the APIs. In this context, we proposed 23 open research questions that can be immediately approached by leveraging the open access model and new programming interfaces.

We will optimize our API interfaces to the needs of the scientific community and zbMATHs data partners in the future.
Depending on the needs of the communities, we will evolve and adapt our data formats.
Moreover, we will advocate for open access publications and permissive licenses for the reuse of scholarly metadata.
We aim to convince publishers to distribute abstracts and references under permissive licenses.
We will also continue to integrate mathematics-related research software and research data besides traditional publications.

\printbibliography%
\end{document}